\input{aipcheck}

\documentclass[
    ,final            % use final for the camera ready runs
  ]
  {aipproc}

\layoutstyle{6x9}
\usepackage{amssymb}
\begin{document}

\title{Effective field theory as the bridge between lattice QCD and
  nuclear physics}

\classification{12.38.-t,12.39.Fe,12.38.Gc
               }
\keywords      {nuclear physics, lattice QCD, effective field
  theory,chiral perturbation theory}

\author{David B. Kaplan}{
  address={Institute for Nuclear Theory, Seattle, WA, 98195-1550, USA}
}

\begin{abstract}
A confluence of theoretical and technological developments are
beginning to make possible contributions to nuclear physics from
lattice QCD. Effective field theory plays a critical role in these
advances.  I give several examples.
\end{abstract}

\maketitle

%%%%%%%%%%%%%%%%%%%%%%%%%%%%%%%%%%%%%%%%%%%%
%% MAINMATTER
%%%%%%%%%%%%%%%%%%%%%%%%%%%%%%%%%%%%%%%%%%%%

\section{Introduction}
While it is unrealistic to expect to see a solution of the structure
of a uranium nucleus from QCD within our lifetimes, it is not
unreasonable to predict that lattice QCD will make significant
contributions to nuclear physics over the next couple of 
decades. Simultaneous progress in computer technology,
computational algorithms, and advances in theory have made it feasible
to begin such a program in lattice nuclear physics now.

The limitation one faces is the computational cost of a realistic
simulation.  L. Giusti presented the following formula at Lattice '06
for the cost (in Tflops-yrs) for generating gauge field configurations
with dynamical
Wilson fermions (http://www.physics.arizona.edu/lattice06/):
\begin{equation}
{\rm Cost} \sim 0.15 \cdot \left[\frac{ \# {\rm
      configs}}{1000}\right]\cdot\left[\frac{ m_q}{20\,{\rm
      MeV}}\right]^{-1} \cdot\left[\frac{V}{32\,{\rm
      fm}^4}\right]^{\frac{5}{4} }\cdot\left[\frac{a}{0.08\,{\rm
        fm}}\right]^{-6} 
\label{cost}
\end{equation}
Here $a$ is the lattice spacing, $m_q$ is the light quark mass, $V$ is
the lattice volume. 

Significant advances in algorithms have occurred in recent
years, with the discovery in the 1990's of how to simulate chiral
fermions \cite{Kaplan:1992bt,Neuberger:1997fp},  and the improvement
of methods for including
light dynamical fermions  (for example, the power of the mass
dependence in the above formula has dropped from   $m_q^{-6}$ to
$m_q^{-1}$ since the development of algorithms in refs.~\cite{Luscher:2003vf,Luscher:2005rx}).  Technological advances
continue unabated, and machines
currently exist operating in the $10^2$ Tflops range, and Pflops
computing will exist before long.
  
Nevertheless, technology plus algorithms do not by themselves add up to advances
in nuclear theory in the near future because of the daunting
computation costs of a realistic simulation.   To avoid the disadvantages of  non-chiral lattice
fermions, such as the Wilson formulation,  one should use
 domain wall or overlap fermions, incurring in the cost another factor of $\sim
100\times$; the correct light quark masses are
$m_u\simeq 2.5$~MeV, and $m_d\simeq 5$~MeV, not 20 MeV;  the box size should be
ample enough to accommodate the hadrons of interest  (the
Compton wavelength of the pion is about $1.4$ fm, while the scattering
length for the deuteron is about 5 fm); and the lattice spacing of the
real world is, of
course, zero.  Finally, the above cost estimate only covers generation of
lattice configurations;  one must also account for the cost of
generating quark propagators, the number of which grows factorially with the number of
quarks involved --- an unfortunate fact highly relevant to the study
of even the smallest nuclei! Lattice QCD studies of a helium nucleus,
for example, require
$6!^2=518,400$ quark propagator contractions.  It is easy to see that
a brute force approach on a Pflops machine
will not  provide useful information about the $\alpha$ particle at
realistic quark masses.  

Effective field theory is the tool that will allow us to extract
useful information from available technology, giving us the ability to
simulate real systems at unrealistic lattice parameters.

In particular, we will have the opportunity to learn about
fundamental properties of matter which are not directly obtainable from
experiment, and which are necessary inputs for reliable nuclear
structure or equation-of-state calculations.  These include an
improved understanding of three-body forces, such as 
in the experimentally inaccessible  $I=3/2$ channel,  and the
interactions between hyperons and 
nucleons. The thesis of this talk is that progress in these directions
will need an intense effort by theorists in order to extract
physically relevant quantities from feasible lattice calculations, and
that the basic tool for this effort will be effective field theory.  

Effective field theory (EFT) in all its forms is basically a perturbative
expansion in the ratio of two length scales.  As such, its validity
requires that there be small ratios to exploit. Chiral perturbation
theory has been the basic EFT exploited in continuum QCD, making use
of the mass
gap between the pion and the heavier hadrons.  In addition,  an
effective field theory for nuclear physics has been 
in the making over the 
past 15 years, which incorporates an additional small ratio, the  QCD  length divided by the $NN$
scattering length. 

What is new when working with {\it lattice} QCD is that there are a host of
additional dimensionful scales which do not exist in the real world, but which can
be profitably exploited.  These include the lattice spacing, the lattice
size, and independently varied  masses  for valence and sea quarks.  
EFT allows one to 
\begin{itemize}
\item extrapolate to smaller quark mass than is feasible to simulate
\item parametrize and correct for  finite lattice spacing errors
\item parametrize and correct for finite volume errors
\item extract physics from ``cheaper'' fermions
\item determine $S$-matrix elements from Euclidean simulations by
  measuring volume dependence of the spectrum
\item extract useful physical quantities from complicated multi-hadron
  systems
\end{itemize}

\section{Uses of chiral perturbation theory}

\subsection{Quark mass extrapolation}
\label{sec:2a}

Chiral perturbation theory is an expansion of the Lagrangian for low
energy QCD about the chiral limit, $m_q=0$.  As such, it is obviously
useful to extrapolate from lattice simulations at somewhat heavy quark
masses, down to realistic quark masses. For this to work, the lattice
quark mass has to be light enough so that the chiral expansion still 
converges. The chiral expansion parameter for mesonic processes is
$m_\pi^2/\Lambda^2$, where 
$\Lambda\sim m_\rho$ is not far from 1~GeV.  A light quark mass of
$20$ MeV, for example, corresponding 
to $m_\pi\sim 325$~MeV, should be within the range of validity
of chiral perturbation theory. I will not dwell on this conventional
and important application of chiral perturbation theory which is
widely familiar (see lectures by S. Sharpe \cite{Sharpe:2006pu} for a comprehensive
introduction to lattice applications of chiral perturbation theory).

\subsection{Lattice spacing extrapolation}
\label{sec:2b}

Another application of chiral perturbation theory is to account for
finite lattice spacing errors. One first matches the lattice action to the
``Symanzik action'' -- a continuum theory with all operators allowed
by the lattice symmetries, suppressed by powers of the lattice spacing
$a$ appropriate to the dimension of the operator.  For example, with
Wilson fermions (which do not possess a chiral symmetry), the leading
operators in the Symanzik action not present in continuum QCD include
\begin{itemize}
\item dimension-3 chiral symmetry violation: $a^{-1} \bar q q$
\item dimension-5 chiral symmetry violation: $a \bar q \sigma_{\mu\nu}
  G_{\mu\nu} q$
\item  dimension-6 Lorentz violation: $a^2 \bar q D_\mu^3 \gamma^\mu q$
\end{itemize}

In order to determine the effects of finite lattice spacing on
low energy hadronic physics, on can then match the Symanzik action
onto a generalized chiral Lagrangian, which includes the effects of
these finite lattice spacing operators \cite{Sharpe:1998xm,Rupak:2002sm,Bar:2003mh}.
The coefficients of these
operators may be determined by making measurements at several lattice
spacings, and then the extrapolation to $a=0$ may be improved.  

This
program is versatile and can be applied to different lattice fermion
formulations. For Wilson fermion  the chiral symmetry violating
operators 
 give rise to the leading $O(a)$ corrections, even when the
dimension-3 operator is 
fine-tuned away.  For staggered fermions, corrections begin at
$O(a^2)$, but the effective theory is
complicated by the presence of additional ``tastes'', with an
approximate $SU(4)$ taste symmetry, broken by finite lattice spacing
operators.  The analysis of the chiral Lagrangian is simplest for
chiral lattice fermions, such as domain wall or overlap fermions,
which automatically avoid the $O(a)$ operators without incurring
spurious fermion tastes. The computational price of dynamical chiral
fermions is about a factor of $100$, which is severe.

\subsection{Partially quenched chiral perturbation theory}

Quark masses appear in two distinct ways in the calculation of a
correlation function in lattice QCD: either in the fermion
determinant, which controls the gauge field configuration one
generates; or in the fermion propagators on sews together in the gauge
field background to compute the desired Green function.  The former is
called the ``sea quark mass'', the latter the ``valence quark mass''.
In the real world they are the same, but in a lattice calculation they
can be different.  By making the valence quark mass light while
keeping the sea quark mass heavier, one obtains
the benefits of chiral symmetry to leading order in the gauge
coupling, without paying the light fermion price in generating the
gauge field configurations.

\begin{figure}
  \includegraphics[height=.25\textheight]{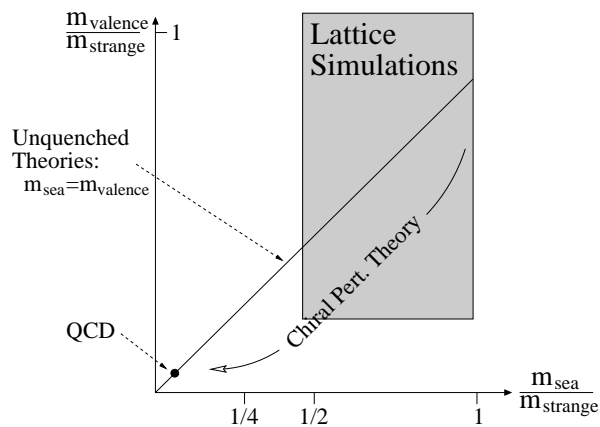}
  \caption{Schematic representation of parameter space in partially
    quenched theories, from \cite{Sharpe:1999kj}.}
\label{mspace}
\end{figure}

One can think of this unphysical partition function arising from 
unphysically heavy $u$ and $d$ ``sea'' quarks with mass $m_S$, plus two flavors of light
``valence'' quarks with mass $m_V$, plus
two flavors of ghosts with mass $m_V$ to cancel the valence quark
contribution to the fermion determinant.  Physical hadrons are then
those made of $VV$ quarks (as well as the strange quark), in the limit
$m_{V} = m_{S}$.  The theory has
additional unphysical mixed states of $VS$ and $SS$ content.  A chiral
Lagrangian can be constructed for this system, which 
contains new operators, but also all the operators of real QCD \cite{Bernard:1993sv,Sharpe:2003vy}.
Provided that $m_S$ is light enough for chiral perturbation
theory to apply, by varying $m_S$ and $m_V$
independently, and by making use of generalized chiral perturbation theory for the
various types of correlation functions one can compute, it is possible to
isolate and measure quantities of interest in QCD, such as the
Gasser-Leutwyler coefficients for $NLO$ chiral perturbation theory
\cite{Sharpe:1999kj}.  See fig.~\ref{mspace} for a picture of the
expanded parameter space.
For an example of how to use the partially quenched method to
determine the up quark mass, see \cite{Cohen:1999kk}; there it was
shown that by computing meson masses in the combination
$(M_{VV}^2+M_{SS}^2 -2 M_{SV}^2)$ one can extract the combination of
Gasser-Leutwyler coefficients $(2L_8-L_5)$ for QCD (where $S$ and $V$
label the two propagators used).

\subsection{Mixed action}

There has been much work done recently with staggered fermions,
employing ``the fourth root trick'' to reduce the number of tastes
from four to one per physical quark flavor.  This results in a
nonlocal theory at finite lattice spacing, and there has been a
controversy about whether or not the resulting lattice theory is in
the same universality class as QCD and is capable of delivering an
approximation to continuum QCD \cite{Creutz:2006wv,Sharpe:2006re}. In addition, the EFT for staggered
fermions is extremely complex and difficult to work with for baryons, due to the
multiplicity of tastes and taste symmetry violating operators.  The
benefits of staggered fermions are
their computational cheapness, and so at least for now, they are
widely employed for mesons (see, for example, \cite{Kronfeld:2005fy}).

It is possible to improve the utility of staggered fermions by
working with mixed actions, where the sea quarks are staggered, while
the valence quarks are domain wall fermions.  This approach benefits
from combining the speed of staggered fermions and the chiral symmetry of
domain wall fermions, and has been
used in recent calculations of $g_A$ at the $\sim 10\%$ level \cite{Edwards:2005ym}, the ratio
$f_\pi/f_K$ to  $\sim 1\%$ \cite{Beane:2006kx}, $\pi-\pi$ scattering
\cite{Beane:2005rj}, $K-\pi$ scattering \cite{Beane:2006gj},  and
two-nucleon properties \cite{Beane:2006mx}. For mixed action
computations one can use the partially quenched chiral perturbation
machinery described above to extract physical results from the lattice
calculations \cite{Bar:2005tu}.

\subsection{Volume dependence}
\label{sec:1c}

An additional handle on QCD provided by the lattice is the ability to
manipulate the volume.  Chiral perturbation theory is useful for
understanding volume dependence of physical quantities, because it is
the lightest modes that are most sensitive to finite volume.  For a
sufficiently small lattice, the zero-momentum pion modes become
collective coordinates corresponding to global 
rotations of the chiral condensate. This occurs when $m\langle\bar q
q\rangle V < 1$, defining the
$\epsilon$-regime  \cite{Gasser:1987zq}.  This
regime was recently cleverly exploited 
to extract information about QCD in the infinite volume continuum for
the $\Delta I=1/2$ rule \cite{Hernandez:2006kz}, and for nucleon
properties \cite{Detmold:2004ap}. Both references make extensive use of
chiral perturbation theory.

\section{Nuclear effective field theory}

EFT will play a critical role in computing nuclear physics properties
on the lattice.  Just as for meson interactions, one would like to
find strategies to measure on the lattice the coefficients of the most relevant
operators which control the interactions of nucleons, and then use
that effective theory to compute nuclear properties.  This program
requires (i) that there be a sensible EFT for the interactions of
nucleons, and  (ii) that one can relate lattice measurements in Euclidean
space to experimentally measurable quantities, such as scattering
lengths.

Nuclear effective theory was pioneered by Weinberg
\cite{Weinberg:1990rz}, first explored in ref.~\cite{Ordonez:1995rz},
and then developed by
many subsequent authors (for a review, see \cite{Bedaque:2002mn}). For
low energy nucleon interactions, the pion may be considered as heavy,
and the pion-less EFT developed in
\cite{vanKolck:1997ut,Kaplan:1998tg,Kaplan:1998we,Chen:1999tn} consists of contact
interactions $\sim C_0 (N^\dagger N)^2 + C_2 (N^\dagger N)(N^\dagger
\nabla^2 N) + O(p^4)\ldots$  The momentum expansion treats $C_{2n}
=O(p^{n-1})$ when renormalized at a scale $\mu=O(p)$, and $C_0$ is
summed to all orders as the leading contribution, while the higher
$C_{2n}$ are inserted perturbatively.  For $p\gtrsim m_\pi/2$, the
pion has to be included explicitly.
Unfortunately, unlike the case for chiral perturbation theory, the power counting
scheme for the effective theory for  nucleons interacting via pions is
somewhat controversial.  The reason 
is in part because $NN$ scattering is nonperturbative, and so the actual scaling
of an
operator does not match its naive dimension, making it difficult
to construct a consistent power counting scheme.  The original
Weinberg scheme suffers inconsistencies, where counterterms appear at
higher orders than the divergences they are supposed to cancel. One
example of this was given in \cite{Kaplan:1996xu} where it was shown
that a quark
mass-dependent counterterm for the non-derivative $NN$ vertex was
required at leading order; a more recent numerical analysis of $NN$ scattering by
the tensor force demonstrated that counterterms are needed at leading
order in all partial waves where the interaction is attractive
\cite{Nogga:2005hy}, even though such counterterms would be subleading in
Weinberg's expansion.  By
working at a fixed and not too large cutoff, this problem can be swept
under the rug, but this procedure  in effect corresponds to constructing a model
for short distance physics  rather than performing a {\it
  bona
fide} EFT calculation. This point of view is not universally accepted,
and for recent contributions on various sides of the controversy see
refs.~\cite{Epelbaum:2006pt,Rho:2006tx,RuizArriola:2006hc}.
The KSW expansion \cite{Kaplan:1998tg,Kaplan:1998we} was offered as an
alternative to Weinberg's expansion, but was found not to converge well for two nucleons in
the $s=1$ channel \cite{Fleming:1999ee}.  I believe that the most
consistent expansion currently available is that of
ref.~\cite{Beane:2001bc}, with generalizations to account for three
nucleon forces \cite{Bedaque:2002yg}. 

There have been many notable successes of the nuclear effective theory, most
remarkably at very low energy.  A nice example is the isolation of the
EFT coupling $L_{1A}$, the analogue of $g_A$ for the axial isovector
two-nucleon current.  By fitting it to data one can compute the
neutrino-deuteron breakup cross section, thereby reducing a major
source of systematic error in the analysis of  data from the Sudbury
Neutrino Observatory \cite{Chen:2002pv}.

In order to study multi-nucleon states on the lattice, one approach is
to compare EFT and numerical results in Euclidean space, and determine
the EFT couplings. Another approach is to compare lattice results for
S-matrix elements with the predictions of the EFT. An important
contribution to the problem of extracting S-matrix 
elements from Euclidean lattice theory  was devised by L\"uscher, who showed
how the volume dependence of the energy for a 2-particle
state in a box yields the scattering lengths \cite{Luscher:1990ux}.  
A nonrelativistic  formulation found in ref.~\cite{Beane:2003da,Beane:2006mx} goes as
follows: The Feynman scattering amplitude for two nucleons has the
form ${\cal A} = (4 \pi/M)/(p\cot \delta(p) - i p)$, where $p=\sqrt{M E}$,
$E$ being the energy in the center of mass. When formulated in a box,
the energy eigenvalues correspond to zeros of $Re[({\cal A})^{-1}]$. So
the energy eigenvalues solve
\begin{equation}
0 = Re[({\cal A})^{-1}]_{\rm box} =   Re[({\cal A})^{-1}]_{L=\infty} +\left(
  Re[({\cal A})^{-1}]_{\rm box}-Re[({\cal A})^{-1}]_{L=\infty}\right)\ .
\end{equation}
The first term on the right  is just proportional to $p\cot\delta$ in an
infinite box.  The second term is the difference between  the bubble
diagram for two nucleons scattering off each other computed in an
infinite box versus  computed in a finite box. This is a finite and
computable function of the box size $L$ and the eigenenergy $E_n$,
and one arrives at the formula
\begin{equation}
p_n\cot\delta(p_n) = \frac{1}{\pi L} S(\eta_n)\ ,\quad
p_n = \sqrt{E_n M}\ ,\quad \eta_n = (p_n L/2\pi)^2\ ,
\end{equation}
\begin{equation}
S(\eta) = \lim_{\Lambda\to\infty} \left[\sum_{|\vec j|<\Lambda}\ 
  \frac{1}{|\vec j|^2 - \eta}-4 \pi\Lambda\right]\ ,
\label{lusch}
\end{equation}
%The function $S(\eta)$ is plotted above.
%\begin{figure}
%  \includegraphics[angle=90, height=.3\textheight]{FS2.eps}
%  \caption{The function S from equ. (4).}
%\label{FS2}
%\end{figure}
where the $\vec j$ are integer triplets. 
By measuring the energy eigenvalues for two nucleons  in a box, one
can then in principle solve the above equation for $\delta(p_n)$. A
pioneering measurement of two nucleon scattering lengths  using this
method 
was performed in ref.~\cite{Beane:2006mx}, where they concluded that
one would need a box of size $5-15$ fm in order to study properties of
the deuteron.  This is a large box, but much smaller than one would
conclude directly for L\"uscher's work, which would seem to indicate
$L\gg a$.

It is important to push two nucleon studies much further, technically and
theoretically.  Using EFT techniques, one should learn how to measure
matrix elements of currents in the two-nucleon state, study
hyperon-nucleon interactions, and prepare the groundwork for the study
of three-nucleon states on Pflops machines.

\section{The frontier}

Lattice QCD can  make substantial and useful contributions by
predicting properties of baryons which are not 
experimentally accessible.  I believe that in the foreseeable future,
useful prediction are feasible in four areas:
\begin{itemize}
\item The masses and couplings of QCD resonances and hybrids;
\item Strangeness physics, such as $\bar K N$, $YY$ and $YN$
  interactions, where $\bar K$ is the anti-kaon,  $Y$ is a hyperon,
  and $N$ is a
  nucleon;
\item Determination of 3-body interactions, such as in the $I=3/2$
  channel;
\item Quark mass dependence of nuclear properties.
\end{itemize}
The first category could make an important contribution to the JLab
experimental program; information the second category could answer
basic questions about dense matter in neutron stars, such as which
hadronic channel is favored when strange quarks first appear, the
$K^-$, $\Lambda$, or $\Sigma^-$?  Accomplishing the third task would
be an important milestone, whereby lattice QCD could inform the
so-called {\it ab initio} nuclear structure calculations and make them
significantly more {\it ab initio}.  The fourth category could be
of interest in understanding how fine-tuned is our world, and could be
important in certain cosmological theories where the quark masses are
dynamically determined quantities \cite{Beane:2002xf}.

All of these projects could be very rewarding, and  will
require an intensive theoretical effort that further develops the
3-nucleon EFT, extends L\"uscher's work to three particles in a box
(where inelastic thresholds could cause problems \cite{Lin:2001ek}), 
and improves the available computational algorithms.  Lattice QCD will
clearly play an important role in the future progress of nuclear
theory, and EFT will be a vital component of the program.
%\begin{equation}
%J_{ion}=A\frac{exp\left[-\frac{E_a}{kT}\right]}{kT}\alpha \label{ionflux}
%\end{equation}
%%%%%%%%%%%%%%%%%%%%%%%%%%%%%%%%%%%%%%%%%%%%%
%%% Sample figure:
%%%
%%% The option [height=...] scales the picture to the given height,
%%% without it it would be printed at its nominal size
%%%%%%%%%%%%%%%%%%%%%%%%%%%%%%%%%%%%%%%%%%%%%

%\begin{figure}
%  \includegraphics[height=.3\textheight]{golfer}
%  \caption{Picture to fixed height}
%\end{figure}

%%%%%%%%%%%%%%%%%%%%%%%%%%%%%%%%%%%%%%%%%%%%%
%%% SAMPLE TABLE
%%%
%%% Shows the use of \tablehead and \tablenote
%%% macros
%%%%%%%%%%%%%%%%%%%%%%%%%%%%%%%%%%%%%%%%%%%%%

%\begin{table}
%\begin{tabular}{lrrrr}
%\hline
%  & \tablehead{1}{r}{b}{Single\\outlet}
%  & \tablehead{1}{r}{b}{Small\tablenote{2-9 retail outlets}\\multiple}
%  & \tablehead{1}{r}{b}{Large\\multiple}
%  & \tablehead{1}{r}{b}{Total}   \\
%\hline
%1982 & 98 & 129 & 620    & 847\\
%1987 & 138 & 176 & 1000  & 1314\\
%1991 & 173 & 248 & 1230  & 1651\\
%1998\tablenote{predicted} & 200 & 300 & 1500  & 2000\\
%\hline
%\end{tabular}
%\caption{Average turnover per shop: by type
%  of retail organisation}
%\label{tab:a}
%\end{table}

\begin{theacknowledgments}
I wish to thank G. Martinelli, C. Sachrajda, S. Sharpe and M. Savage
for useful conversations, and to the organizers of QCHSVII for their hospitality.
This work was supported by the US Department of Energy grant DE-FG02-00ER41132.
\end{theacknowledgments}



\bibliographystyle{aipproc}   % if natbib is available

\bibliography{David_Kaplan}

\IfFileExists{\jobname.bbl}{}
 {\typeout{}
  \typeout{******************************************}
  \typeout{** Please run "bibtex \jobname" to obtain}
  \typeout{** the bibliography and then re-run LaTeX}
  \typeout{** twice to fix the references!}
  \typeout{******************************************}
  \typeout{}
 }

\end{document}